# Study of sedimentation of non-cohesive particles via CFD–DEM simulations

S.L. Xu[1]; Rui Sun[2]; Y.Q. Cai[3]; H.L. Sun[4*]

1. S.L. Xu, Ph.D. Candidate, Research Center of Coastal and Urban Geotechnical Engineering, College of Civil Engineering and Architecture, Zhejiang University, Anzhong Bldg A404, Hangzhou 310058, China, E-mail: xuxiaoshan7@126.com

2. Rui Sun, Ph.D. Candidate, Department of Aerospace and Ocean Engineering, Virginia Tech, 311-A Randolph Hall (0203), Blacksburg, VA 24061, USA, E-mail: sunrui@vt.edu

3. Y.Q. Cai, Professor, Research Center of Coastal and Urban Geotechnical Engineering, College of Civil Engineering and Architecture, Zhejiang University, Anzhong Bldg A404, Hangzhou 310058, China, E-mail: caiyq@zju.edu.cn

4*. H.L. Sun, Corresponding author (✉), Associate Professor, Institute of Disaster Prevention, College of Civil Engineering and Architecture, Zhejiang University, Anzhong Bldg B832, E-mail: sunhonglei@zju.edu.cn, Telephone: +86 13758217424

# Study of sedimentation of non-cohesive particles via CFD–DEM simulations

Shan-Lin Xu[1]; Rui Sun[2]; Yuan-qiang Cai[1]; Hong-lei Sun[3]

**Abstract**

The sedimentation process of granular materials exists ubiquitously in nature and many fields which involve the solid–liquid separation. This paper employs the coupled computational fluid dynamics and discrete element method (CFD–DEM) to investigate the sedimentation process of non-cohesive particles, including the hindered settling stage and the deposition stage. Firstly, the coupled CFD–DEM model for sedimentation is validated by the hindered settling velocity at different solid volume concentrations of suspension $\phi_0$, i.e., $\phi_0 = 0.05$~$0.6$. Two typical modes of sedimentation are also presented by the concentration profiles and the equal-concentration lines. Then, the comparisons between mono- and poly-dispersed particle system are detailed. In the sedimentation of the poly-dispersed particle system, the segregation phenomenon is simulated. Furthermore, this segregation effect reduces with the increase of the initial solid concentration of suspension. From the simulations, the contact force between every pair of particles can be obtained, hence we demonstrate the "effective stress principle" from the view of the particle contact force by giving the correspondence between the particle contact force and the "effective stress", which is a critical concept of soil mechanics. We also demonstrate the effective stress principle from the view of the contact force acting on particles. Moreover, the deposition stage can be simulated by CFD–DEM method, therefore the solid concentrations of sediment bed $\phi_{max}$ on different conditions are studied. Based on the simulation results of $\phi_{max}$ and the theory of sedimentation, this paper also discusses a method to calculate the critical time when sedimentation ends of two typical modes of sedimentation.

**Keywords:** Sedimentation; Hindered settling; CFD–DEM; Effective stress; Sediment bed density; Critical time

# 1 Introduction

The sedimentation process of suspended particles in the fluid is of great practical importance, which can be found in many fields, such as the deposition of natural sediments, the numerous solid–liquid separation phenomena in chemical, mining, food and pharmaceutical engineering [1-4]. In a wide variety of particle–fluid system, the non-cohesive particles go through the hindered settling process and the deposition process. During the hindered settling process, particles fall at a constant velocity due to the balance of forces. Moreover, the falling velocity of the particles will reduce with the increase of local particle concentration because of the increase of collisions and interferences between particles [5,6]. Then, with the deposition of particles, the settling velocity of particles decreases to zero, and the sediment bed is gradually formed.

✉ Hong-lei Sun
E-mail: sunhonglei@zju.edu.cn
Shan-lin Xu
E-mail: xuxiaoshan7@126.com
Rui Sun
E-mail: sunrui@vt.edu
Yuan-Qiang Cai
E-mail: caiyq@zju.edu.cn
[1] Research Center of Coastal and Urban Geotechnical Engineering, College of Civil Engineering and Architecture, Zhejiang University, Anzhong Bldg A404, Hangzhou 310058, China
[2] Department of Aerospace and Ocean Engineering, Virginia Tech, 311-A Randolph Hall (0203), Blacksburg, VA 24061, USA
[3] Institute of Disaster Prevention Engineering, College of Civil Engineering and Architecture, Zhejiang University, Anzhong Bldg B832, Hangzhou 310058, China

The sedimentation process has been studied theoretically and experimentally for decades. The theory of sedimentation was firstly studied by Kynch [5]. He assumed that the settling velocity was only determined by the local particle concentration and proposed a mass continuity equation to describe the concentration's variations. Based on Kynch's theory, Bustos and Concha detailed the settling behavior with the method of characteristics line and presented the mathematical analysis on the different modes of sedimentation [7,8]. Moreover, Kranenburg [9] and Dankers [6] applied and extended this theory on the modeling of the hindered settling of the mud flocs. There are also numerous experimental studies on sedimentation process. The majority of these studies focused on investigating the settling velocity and the velocity–concentration relationship [1,11-13]. Among these studies, Richardson's empirical formula [1] about hindered settling velocity was widely accepted and utilized in many particle–fluid systems.

In recent twenty years, some numerical methods have been proposed to study the sedimentation process of various suspensions, such as the finite differencing Navier-Stokes solver which can be used to simulate the bulk sedimentation process [14,15]; the Lattice Boltzmann method (LBM) utilized to study the differential settling process [16,17]; the two fluid models (TFM) [4,18] proposed to simulate some specific mixtures, and the CFD–DEM coupling method to simulate the batch sedimentation [19]. Among these numerical methods, the coupled CFD–DEM method is attractive to simulate the particle–fluid flow because of its superior computational convenience as compared to LBM and capability to capture the particle physics as compared to TFM[20]. In this paper, we employed the CFD–DEM method to simulate the sedimentation process. The effective hindered settling velocity of the suspension at such wide range of solid concentration, $\phi_0 = 0.05$~$0.6$, was firstly simulated and compared with the results calculated by Richardson's empirical formula. The agreement between these two methods demonstrated the validity and ability of our simulation model. We also used this model to study the two typical modes of mono-dispersed particle system sedimentation and the poly-dispersed system sedimentation. Thanks to the DEM simulation, we could obtain the contact force between any contacted particles. So for the first time, we demonstrated the relationship between the micro-scale contact stress among any contacted sand particles and the critical macro-scale concept "effective stress" defined in soil mechanics. In the end, we discussed a method to obtain the sedimentation time, which was rarely studied in previous research and of great importance in the practical designs.

The paper is organized as follows. Section 2 reviews the theory of hindered settling and presents the methodology of CFD–DEM model. Section 3 shows the simulation results. In this part, the settling velocity–concentration relationship, the two typical sedimentation modes of mono-dispersed particle system, and the segregation phenomenon of poly-dispersed particle system are both simulated. Then, the variations of fluid pressure, the particle–particle contact force, and the relationship between the mean contact stress and the "effective stress" is studied. Moreover, the solid concentrations of sediment beds on different conditions are obtained from simulations. Section 4 discusses a method to estimate the sedimentation time when sedimentation ends based on the theory of the sedimentation. Section 5 concludes the paper.

## 2 Theory and methodology

**2.1 Overview of the theory of hindered settling**

When a single particle settles in still fluid, it has a specific falling velocity determined by the physical properties of particle and fluid. With the increase in the number of particles, the settling velocity of the group of particles reduces because of particles hindering each other and the effect of return flow caused by the downward settling of particles [6]. This process is called the hindered settling. Kynch studied this process and proposed that the falling speed of mono-dispersed particle system is determined by the local particle concentration only [5,6,9]:

$$V_s = V_{(s,0)} f(\phi), \tag{1}$$

$$S(\phi) = V_s \phi = V_{(s,0)} \phi f(\phi), \tag{2}$$

where $V_{(s,0)}$ is the terminal velocity of a single particle and $V_s$ is the effective falling velocity of the superficial suspension; $\phi$ means the solid volume concentration of suspensions; $S$ is the particle flux which represents the volume of particles crossing a horizontal section per unit area per unit of time. To describe the settling process, the one-dimensional mass conversation equation of suspensions is written as [9]:

$$\frac{\partial \phi}{\partial t} + \frac{\partial S(\phi)}{\partial z} = 0, \tag{3}$$

Eq. (2) could be substituted to the Eq. (3):

$$\frac{\partial \phi}{\partial t} + F(\phi) \frac{\partial \phi}{\partial z} = 0, \tag{4}$$

where

$$F(\phi) = S'(\phi) = V_{(s,0)} \frac{d(\phi f(\phi))}{d\phi}. \tag{5}$$

Equation (4) can be solved by using characteristic lines, i.e. lines of equal concentration $\phi$, in the (*z*, *t*) plane [6]. These lines are given by:

$$\frac{dz}{dt} = F(\phi), \tag{6}$$

$$z(t) = z_0(\phi) + F(\phi)t. \tag{7}$$

Where *z* is the vertical coordinate and $z_0(\phi)$ represents the initial height of specific concentration when $t = 0$. Since $F(\phi)$ is constant along the characteristic lines, so these lines are straight in the (*z*, *t*) plane. The process of sedimentation can be described as a phenomenon of wave propagation. The shock waves or the rarefaction waves will occur depending on whether the characteristic lines intersect. Where characteristic lines intersect, a sudden change in concentration, also called the shock (interface) would happen. The propagation speed of the shock wave is determined by the Rankine Hugoniot jump condition [7,8]:

$$U_{\text{shock}}(\phi^+, \phi^-) = \frac{S(\phi^+) - S(\phi^-)}{\phi^+ - \phi^-}. \tag{8}$$

Moreover, this shock occurs when Oleinik's jump entropy condition is fulfilled [8] (see Eq. (9)), which means the interface stability exists.

$$S'(\phi^+) \geq U_{\text{shock}}(\phi^+, \phi^-) \geq S'(\phi^-), \tag{9}$$

where $\phi^+$, $\phi^-$ is the solid volume concentration just above and below the shock, respectively. If Eq. (9) is not satisfied, characteristic lines do not intersect, and a gradual transition of concentration (rarefaction wave) will happen.

From above theory, the hindered settling process is described by the conservation equation of solid volume concentration without taking into account the various forces of the particle–fluid system. In our simulations, the force models of the particle–particle and the particle–fluid are utilized to simulate the sedimentation process, and the simulation results are validated by comparing with the above-mentioned hindered settling theory.

## 2.2 Coupled CFD–DEM approach

What we employ to simulate the hindered settling process is sediFoam, a hybrid CFD–DEM (Computational Fluid Dynamics–Discrete Element Method) solver for particle-laden flows based on two open-source codes: OpenFOAM, an CFD toolbox developed primarily by OpenCFD Ltd, and LAMMPS, a highly efficient molecular dynamics solver distributed by Sandia National Laboratories. The coupled solver is developed by Sun and Xiao [21], and the source code is available at https://github.com/xiaoh/sediFoam. The detailed algorithms of this solver are published in [22,23] and it has been rigorously verified and validated in a wide range of application areas, such as the fluidized bed [24], sediment transport [21] and sand dune migration [25]. Here, a block flow diagram of sediFoam is shown in Fig. 1.

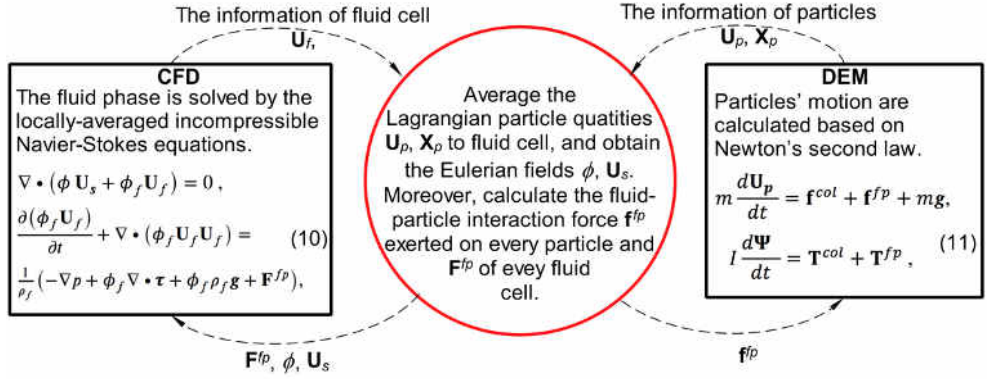

**Fig. 1 The block flow diagram of sediFoam**

The fluid phase is solved in CFD program by the locally-averaged incompressible Navier–Stokes equations [22,26]:

$$\nabla \cdot \left( \phi_s \mathbf{U}_s + \phi_f \mathbf{U}_f \right) = 0,$$

$$\frac{\partial \left( \phi_f \mathbf{U}_f \right)}{\partial t} + \nabla \cdot \left( \phi_f \mathbf{U}_f \mathbf{U}_f \right) = \frac{1}{\rho_f} \left( -\nabla p + \phi_f \nabla \cdot \tau + \phi_f \rho_f \mathbf{g} + \mathbf{F}^{fp} \right), \tag{10}$$

where $\phi_s$ is the solid volume fraction of a fluid cell; $\phi_f = 1 - \phi_s$ is the fluid volume fraction; $\mathbf{U}_f$ is the fluid velocity, $\mathbf{U}_s$ is the solid/particle-phase velocity, which is obtained by averaging the Lagrangian particle velocities in the Eulerian fields. The averaging process was introduced in [21,22]. The right terms of the momentum conservation equation are the pressure gradient $\nabla p$, the divergence of the stress tensor $\tau$, gravity g, and fluid–particle interaction forces $F^{fp}$, which is the averaged interaction force from every individual particle in a fluid cell. The Eulerian fields $\phi$, $\mathbf{U}_s$ and $\mathbf{F}^{fp}$ are all obtained by averaging the information of Lagrangian particles.

Particles' motions are tracked in DEM module based on Newton's second law by DEM [27]:

$$m \frac{d\mathbf{U}_p}{dt} = \mathbf{f}^{col} + \mathbf{f}^{fp} + m\boldsymbol{g},$$

$$I \frac{d\boldsymbol{\Psi}}{dt} = \mathbf{T}^{col} + \mathbf{T}^{fp}, \tag{11}$$

where $\mathbf{U}_p$ and $\Psi$ is the velocity and angular velocity of particle ($\mathbf{X}_p$ in Fig. 1 is the coordinates of the particle); $I$ represents the angular moment of inertia of particles; $\mathbf{f}^{fp}$ and $\mathbf{f}^{col}$ denotes the fluid–particle interaction force and the contact force exerted on every particle; $\mathbf{T}^{fp}$ **and** $\mathbf{T}^{col}$ represents the torque due to the fluid–particle interactions and collisions between particles, respectively. A linear spring–dashpot model is utilized to simulate the particle-particle and particle-wall collision during the settling. The normal and tangential interaction forces $\mathbf{f}^{col}_{n,ij}$ and $\mathbf{f}^{col}_{t,ij}$ between two contacted particles $i$ and $j$ is [28]:

$$\mathbf{f}_{n,ij}^{\text{col}} = k_n \delta_{ij} \mathbf{n}_{ij} - \gamma_n m_{\text{eff}} \mathbf{V}_{n,ij}$$
$$\mathbf{f}_{t,ij}^{\text{col}} = -k_t \Delta s_t - \gamma_t m_{\text{eff}} \mathbf{V}_{t,ij} \tag{12}$$

where $k_n$ and $k_t$ are the elastic stiffness constant for normal and tangential contact, $\gamma_n$ and $\gamma_t$ are the viscoelastic damping coefficient for normal and tangential collisions, $\delta_{ij}$ is the overlap distance of two contacted particles $i, j$, $\mathbf{r}_{ij}$ is direction vector connecting the centers of two particles, $\mathbf{n}_{ij}$ is the unit vector of $\mathbf{r}_{ij}$. $m_{\text{eff}} = m_i m_j/(m_i + m_j)$ is the effective mass of the spheres with masses $m_i$ and $m_j$, $\mathbf{V}_{n,ij}$ and $\mathbf{V}_{t,ij}$ are the normal and tangential component of the relative velocity of the two particles, $\Delta s_t$ is the tangential displacement vector between the two spherical particles which is truncated to satisfy a frictional yield criterion. This yield criterion is characterized by frictional coefficient $\mu$. The tangential force $\mathbf{f}_{t,ij}^{\text{col}}$ between two particles grows according to a tangential spring and dash-pot model until $\mathbf{f}_{t,ij}^{\text{col}}/\mathbf{f}_{n,ij}^{\text{col}} = \mu$ and is then held at $\mathbf{f}_{t,ij}^{\text{col}} = \mu \mathbf{f}_{n,ij}^{\text{col}}$ until the particles lose contact. The total force $\mathbf{f}_i^{\text{col}}$ and torque $\mathbf{T}_i^{\text{col}}$ acting on particle $i$ is [28]:

$$\mathbf{f}_i^{\text{col}} = \sum_j (\mathbf{f}_{n,ij}^{\text{col}} + \mathbf{f}_{t,ij}^{\text{col}})$$
$$\mathbf{T}_i^{\text{col}} = \frac{1}{2} \sum_j \mathbf{r}_{ij} \times \mathbf{f}_{t,ij}^{\text{col}} \tag{13}$$

The exchange of information of fluid and particles between the CFD part and DEM part is shown in Fig. 1. The formula of the drag force $\mathbf{f}_d$ employed in our simulations is corrected experimentally with considering the hindered settling effect. It bases on numerous experiment results of the effective falling velocity of the suspensions with different concentration, and could be formulated as [29].

$$\mathbf{f}_{d,i} = \frac{\pi d_{p,i}^3}{6} \frac{1}{\phi_{f,i} \phi_i} \beta_i (\mathbf{U}_{p,i} - \mathbf{U}_{f,i}), \tag{14}$$

where $\mathbf{f}_{d,i}$ represents the drag force exerted on every particle, $\mathbf{U}_{f,i}$ is the fluid velocity interpolated at the location of a particle $i$, and $\beta_i$ is the drag correlation coefficient which accounts for the presence of other particles.

$$\beta_i = \frac{3}{4} \frac{C_{d,i}}{V_{r,i}^2} \frac{|\mathbf{U}_{p,i} - \mathbf{U}_{f,i}|}{d_{p,i}} \phi_{f,i} \phi_i, \text{ with } C_{d,i} = (0.63 + 4.8 \sqrt{V_{r,i}/\text{Re}_{p,i}})^2, \tag{15}$$

where the particle Reynolds number $\text{Re}_{p,i}$ is defined as: $\text{Re}_{p,i} = d_{p,i} |\mathbf{U}_{p,i} - \mathbf{U}_{f,i}|/\nu$, $\nu$ is the kinetic viscosity of the fluid; $V_r$ is the ratio of the terminal velocity of a group of particles to the terminal velocity of a single particle $\mathbf{U}_{p,i}$, obtained from the following correlation [11,29]:

$$V_{r,i} = 0.5 \left( A_{1,i} - 0.06 \text{Re}_{p,i} + \sqrt{(0.06 \text{Re}_{p,i})^2 + 0.12 \text{Re}_{p,i}(2 A_{2,i} - A_{1,i}) + A_{1,i}^2} \right), \tag{16}$$

with

$$A_{1,i} = \phi_{f,i}^{4.14}$$
$$A_{2,i} = \begin{cases} 0.8 \phi_{f,i}^{1.28}, & \text{if } \phi_{f,i} \leq 0.85 \\ \phi_{f,i}^{2.65}, & \text{if } \phi_{f,i} > 0.85 \end{cases} \tag{17}$$

The fluid–particle force considered in this paper includes the drag force, the buoyancy and the fluid pressure gradient force. There are also some other possible fluid–particle force terms, such as virtual mass, lift force, and Basset force [30]. The virtual mass may play an important role when particles have accelerations, lift forces including Saffman lift forces and Magnus lift forces account for the lift or upward force due to particle rotations [31]. In our simulations of hindered settling process, particles mainly fall in the vertical direction with rather little acceleration, in which the effect of lift forces and virtual mass is unobvious. Moreover, this work aims to further the comprehension of the sedimentation process on the different conditions, rather than a completely accurate model. Therefore, the virtual mass forces and lift forces are not considered here.

**2.3 Numerical set-up**

There are three different numerical cases performed in the present work. Case 1 aims to simulate the mono-dispersed spherical sand particles settling in quiescent water, Case 2 simulates the ballotini particles settling in silicon, and Case 3 simulates the poly-dispersed sand particles settling in quiescent water. The physical properties of the fluid flow and the sediment particles are detailed in Table 1. To investigate the settling velocity at different solid concentration, the initial solid volume concentrations of suspensions vary from 0.05 to 0.6 in Case 1 and Case 2, with the initial locations of particles being randomly generated in the computational domain. The geometry and the mesh resolution of the simulation domain are shown in Fig. 2 and Tab. 1, in which the *x*-, *y*- and *z*- coordinates represent the length, width, and height direction, respectively. The size of the CFD mesh is 2.5 times of the particle diameter $d_p$ in Case 1 and 2, which is in the moderate range $[2d_p, 4d_p]$ of many validated numerical cases mentioned in [23,24]. In *x*- and *y*- directions, the periodic boundary conditions are used in CFD models and DEM models. In *z*-direction, two types of "wall boundary" are employed for CFD simulations. On the bottom, the fluid velocity is set as zero due to the boundary layer; and on the top, the slip-wall is used representing the zero-gradient fluid velocity. In addition, a frictional wall boundary is employed to bound the particles along the z-direction in DEM models.

A linear spring–dashpot model is utilized to simulate the particle-particle collisions and the particle-wall collisions during the settling process. The particle time step (DEM time step) should be large enough to avoid the particle interpenetration, yet not so large as to induce too much calculation consumption, which typically set as the value of $t_c/50$ [28]. Quantity $t_c$ represents the particle contact time, and its value depends on $k_n$, $\gamma_n$, and the masses of two contacted particles. The fluid time step (CFD time step) is a fraction 0.1 of the maximum allowable time step from the Courant–Friedrichs–Lewy condition [32]. The detailed parameters of numerical simulations are shown in Table 1.

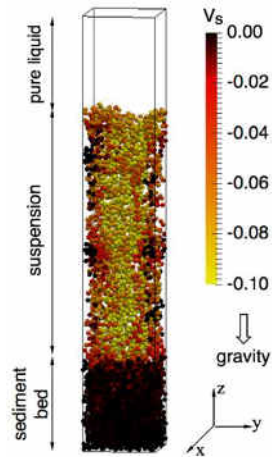

**Fig. 2 Layout of the simulation of sedimentation. During the sedimentation, there are three parts: (1) pure fluid; (2) suspension; (3) sediment bed. $V_s$ represents the effective settling velocity of a group of particles**

Tab. 1 The parameters of numerical simulations

|  | Case 1 (mono-sand in water) | Case 2 (mono-ballotini in silicon) | Case 3 (poly-sand in water) |
|---|---|---|---|
| Geometry |  |  |  |
| size of domain (m) ($x \times y \times z$) | 0.0225×0.0225×0.135 (15 $d_p$× 15$d_p$ ×90 $d_p$) | 0.015×0.015×0.06 (15 $d_p$× 15$d_p$ ×60 $d_p$) | 0.0225×0.0225×0.135 (15 $d_p$× 15$d_p$ ×90 $d_p$) |
| mesh resolutions | 6×6×36 | 6×6×24 | 6×6×36 |

| Particle properties | | | |
|---|---|---|---|
| diameter $d_p$ (mm) | 1.5 | 1.0 | [0.006, 0.27] |
| density $\rho_p$ (kg/m$^3$) | 2650 | 2976 | 2650 |
| stiffness constant $k_n/k_t$ (N/m) | 5000/1428 | 500/143 | 5000/1428 |
| damping constant $\gamma_n/\gamma_t$ (N/ms) | 54000/27000 | 28650/14325 | 54000/27000 |
| friction coefficient $\mu$ | 0.4 [0, 0.4] | 0.02 [0, 0.4] | 0.4 |
| volume fraction | 0.05,0.1,0.2,0.3,0.4,0.5,0.6 | 0.05,0.1,0.2,0.3,0.4,0.5,0.6 | 0.05,0.1,0.2,0.3,0.4 |
| Fluid properties | | | |
| density $\rho_f$ (kg/m$^3$) | 1000 | 1221 | 1000 |
| kinetic viscosity $\nu$ (×10$^{-6}$ m$^2$/s) | 1.01 | 92.46 | 1.01 |

**Note:** The friction coefficient μ of Case 1 and 2 is 0.4 and 0.02, respectively. They both varies from 0 to 0.4 to discuss the influence of friction force between the particles on the settling process and the deposited sediment bed in Sec. 3.6. In Case 3, the diameters of poly-dispersed sand particles range from 0.006 mm to 0.27 mm; mass median diameter of poly-dispersed sand particles $d_{50}$ = 1.5 mm; geometric standard deviation $\sqrt{d_{84}/d_{16}}$ = 1.22.

## 3 Simulation results

### 3.1 The settling curves

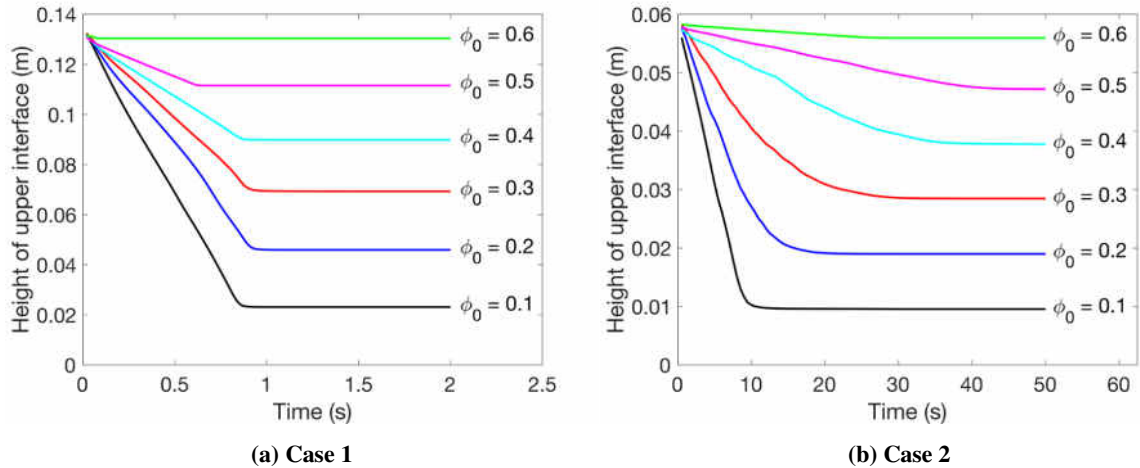

(a) Case 1      (b) Case 2

**Fig. 3** The settling curves obtained in (a) Case 1 and (b) Case 2; The settling curves indicate the location of superficial suspension

In our simulations, the height of top layer particles (2 % of the total particles) was averaged to plot the settling curves as shown in Figs. 3 (a) and (b). We find that the settling curves consist of three parts: (1) an initial slope representing the hindered settling process, during which the particles sustain a fixed averaged falling velocities; (2) a gently curved line indicating the decrease of particle settling velocity corresponding to the deposition stage; and (3) a final horizontal line denoting the end of settling. The first part of the curve is similar to the experimental hindered settling curve of the bitumen froth and mud, as shown in [2,6,33]. The gradient of the first part represents the effective falling velocity, which will be compared with the Richardson's empirical formula in Sec. 3.2. The curved parts of Case 1 and 2 are little different, which represent two typical modes of sedimentation, and will be interpreted in Sec. 3.3. The third part of the settling curve is a horizontal line, not a slight sloping line corresponding to the compaction process during the deposition of cohesive particles (e.g., the consolidating process of the silt/clay in the civil engineering). The present model utilized here is to investigate the non-cohesive particles, so the compaction process is not observed here.

### 3.2 The validation of effective settling velocity

The relationship between the effective velocity and the concentration is crucial on modeling the hindered settling process [34]. Therefore, the effective settling velocities at different initial solid volume concentrations are firstly acquired so as to validate the present numerical model. A widely accepted empirical formula about velocity–concentration relationship was proposed by Richardson [1] based on abundant settling experiments of non-cohesive spherical particles, as shown in Eq. (18).

$$\log V_s = n \log \phi_f + \log V_{(s,0)} = n \log(1 - \phi) + \log V_{(s,0)}, \tag{18}$$

where $\phi_f$ and $\phi$ is the liquid and solid volume concentration of suspension. The values of the empirical parameter $n$ for different flow regimes are as below [1]:

$$n = 4.65 + 19.5 \frac{d_p}{D} \qquad (\text{Re}_{p,0} < 0.2), \tag{19}$$

$$n = \left(4.35 + 17.5 \frac{d_p}{D}\right) Re_{p,0}^{-0.03} \quad (0.2 < \text{Re}_{p,0} < 1), \tag{20}$$

$$n = \left(4.45 + 18 \frac{d_p}{D}\right) Re_{p,0}^{-0.1} \quad (1 < \text{Re}_{p,0} < 200), \tag{21}$$

$$n = 4.45 Re_{p,0}^{-0.1} \qquad (200 < \text{Re}_{p,0} < 500), \tag{22}$$

where $\text{Re}_{p,0}$ is Reynolds number with regard to a particle, defined as [1]:

$$\text{Re}_{p,0} = \frac{V_{(s,0)} d_p}{\nu}. \tag{23}$$

$d_p/D$ is the particle diameter dividing the size of vessel which represents the wall effect. However, such strong wall effects have not been confirmed subsequently, and it's still doubtful if the factor $d_p/D$ has such significant influence [35]. Di Felice recommended $n = 4.65$ if neglecting the wall effect when $\text{Re}_{p,0} < 0.2$.

We compared the settling velocities obtained from numerical simulations and those predicted by Richardson's formula as shown in Fig. 4. In Case 1: terminal velocity of a particle $V_{s,0}$ is 0.1980 m/s, particle Reynolds number $\text{Re}_{p,0}$ is 297.1, so $n$ is calculated by Eq. (22), of which the value is 2.518. In Case 2: $V_{s,0} = 8.174 \times 10^{-3}$ m/s, $\text{Re}_{p,0} = 8.840 \times 10^{-2}$; Since the lateral boundaries are periodic in the simulations, the wall effect is not considered when using Richardson's empirical formula, so the value of $n$ is taken as 4.65.

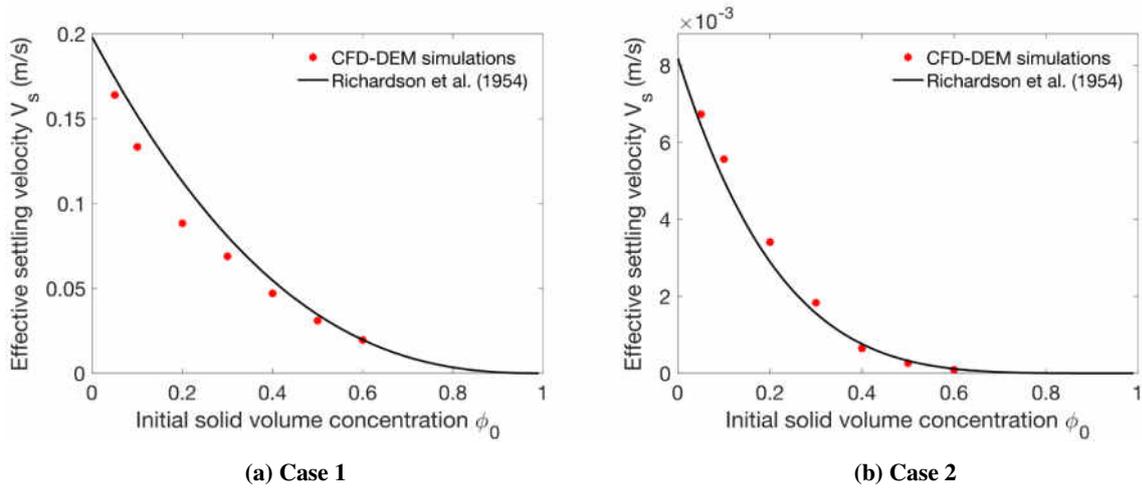

(a) Case 1    (b) Case 2

**Fig. 4 Comparison of effective settling velocities obtained from simulations and those calculated by Richardson's empirical formula in (a) Case 1 and (b) Case 2**

From Fig. 4, the terminal velocities obtained from CFD–DEM are in good agreement with that calculated by the empirical formula. The mutual deviations of velocity at different concentrations obtained by these two methods in Case 1, and Case 2, are in the range [0.4%, 21.7%] and [4%, 20.9%], respectively. The mean deviation at different concentrations is 11.3% in Case 1 and 15% in Case 2. Compared with the errors in the experimental measurement, such discrepancy is acceptable. These agreements demonstrate that the CFD–DEM simulations are able to obtain the settling velocities on different initial conditions in different regime of fluid. Moreover, the drag model and the particle contact model we employed here are suitable to simulate the hindered settling process. According to Kynch [5], the settling velocity of particles decreases to zero when solid concentration increases to its maximum packing value $\phi_{max}$. However, when using the Richardson empirical formula to calculate the settling velocities of $\phi_{max}$, the velocity is a minimum value but not zero. We will discuss how to process this discrepancy in Section 4.

### 3.3 The sedimentation modes of mono-dispersed particle system

Kynch classified the different modes of sedimentation by using the $S - \phi$ curves, and Bustos [7,8], furthermore, gave the rigorous mathematical analysis of the different modes of sedimentation. Among these different modes of sedimentation, we focus on two sedimentation mode depending on whether the lower interface stability exists. These two sedimentation modes can be illustrated by the vertical concentration profiles and the characteristic lines from the simulation results, as shown in Figs. 5 and 6. The occurrence of these two sedimentation modes can be determined by the Oleinik's jump entropy condition (see Eq. (9)).

When Eq. (9) is satisfied, the sedimentation with two obvious interfaces (shocks) occurs, such as in the all series of simulations in Case 1 and several simulations in Case 2 when $\phi_0 = 0.05$, 0.1, 0.5, and 0.6. The vertical concentration profiles and the equal-concentration lines (characteristic lines) of this mode sedimentation obtained from simulations are shown in Fig. 5 (Case 1 when $\phi_0 = 0.2$). From Fig. 5 (a), there are two apparent relative flat stages occurring during the settling process, e.g. from 0 to 0.7 s. These flat stages corresponding to the abrupt changes in concentration mean that the two interfaces: (1) pure water–suspension interface; (2) suspension–sediment interface both develop during the sedimentation. Moreover, these two interfaces divide the mixtures into three parts along the vertical direction, i.e., the pure water, of which $\phi = 0$; the suspensions, of which $\phi = \phi_0$; the sediment bed consisting of deposited particles, of which the concentration is termed as $\phi_{max}$ here. The simulation results of the equal-concentration points are shown in Fig. 5 (b) with the regressed lines. It is found that the vertical coordinate $z$ of the equal-concentration lines almost varies linearly with time, which is the same as the theoretical linear form of characteristic line in Eq. (7). Those equal-concentration lines of which the concentration $\phi \leq \phi_0$ depict the sudden change in concentration from 0 to $\phi_0$ due to the particles falling. Those equal-concentration lines of high concentrations ($\phi > \phi_0$), such as $\phi_0 = 0.3, 0.4, 0.5, 0.6$ are overlapping. This form indicates that the solid volume concentration $\phi$ near the suspension–sediment bed interface changes abruptly from $\phi_0$ to $\phi_{max}$, and the sediment bed concentration reach to the maximum value $\phi_{max}$ instantly once particles accumulate on the bottom of the simulation domain. The area between the high and low equal-concentration lines represents the region of suspensions.

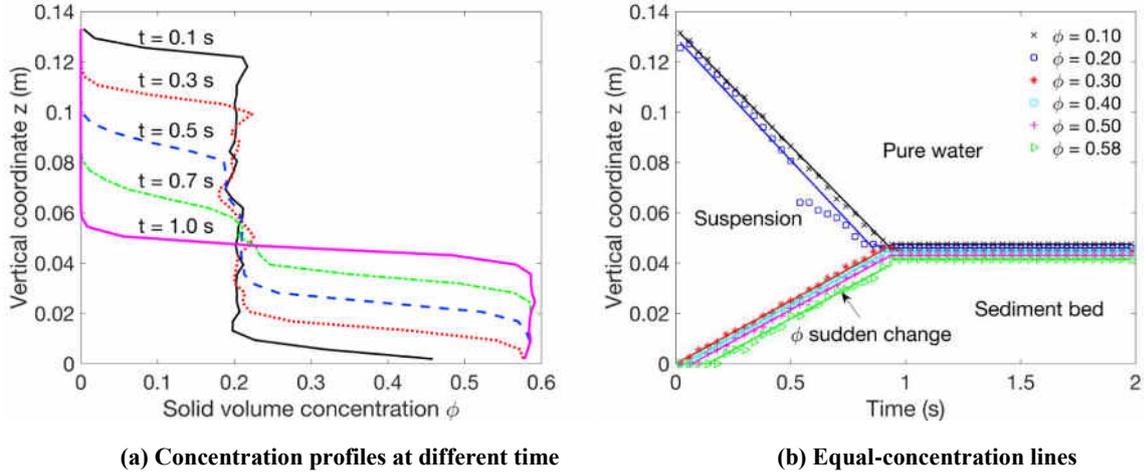

(a) Concentration profiles at different time    (b) Equal-concentration lines

Fig. 5 The (a) concentration profiles and (b) equal-concentration lines obtained in Case 1 when the initial concentration $\phi_0 = 0.2$

If Eq. (9) is unsatisfied, another mode of sedimentation with the only upper interface (pure liquid–suspension interface) occurs, such as in the numerical simulations of Case 2 when $\phi_0 = 0.2, 0.3, 0.4$. This mode of sedimentation is observed in Fig. 6 (Case 2 when $\phi_0 = 0.2$). The upper flat stage is clearly captured. However, the lower flat stage does not arise in the sedimentation process (see Fig. 6 (a)). It means that the concentration near the superficial sediment bed changes gradually from $\phi_0$ to $\phi_{max}$, so there is no obvious interface between the sediment bed and suspension. Moreover, the configuration of equal-concentration lines of which $\phi > \phi_0$ in Case 2 (see Fig. 6 (b)) is different from that in Case 1 (see Fig. 5 (b)). The equal-concentration lines in Case 2 look like a fan radiating from the origin, which are termed as rarefaction wave [7] and observed in settling experiments [6]. These lines illustrate that the concentration of sediment bed gradually increases to the maximum value. Thereby, there is a concentration transition zone between the sediment bed of which $\phi = \phi_{max}$ and the suspension of which $\phi = \phi_0$.

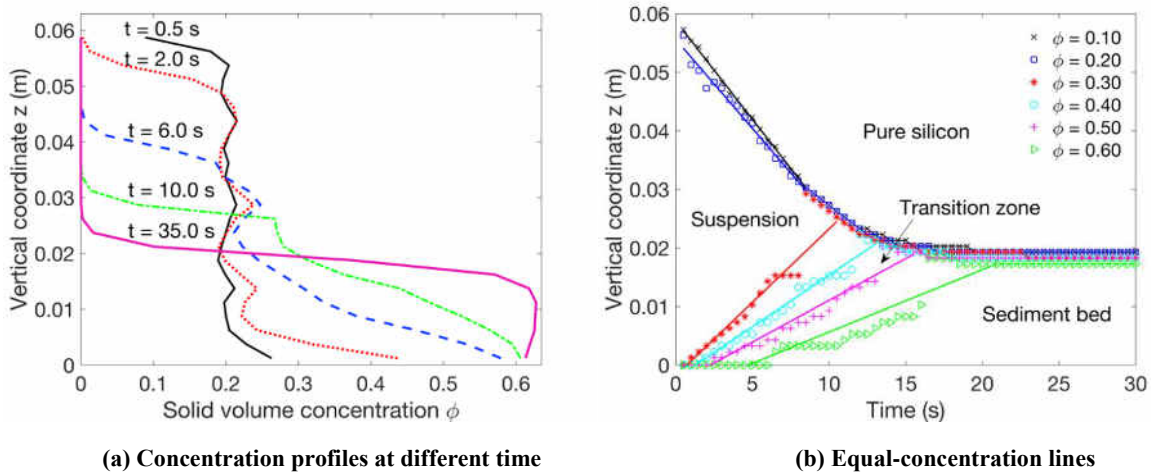

(a) Concentration profiles at different time    (b) Equal-concentration lines

Fig. 6 The (a) concentration profiles and (b) equal-concentration lines obtained in Case 2 when the initial concentration $\phi_0 = 0.2$

The first mode of sedimentation as shown in Fig. 5 is the same as the MS-I sedimentation described by Bustos and Concha [7,8]. A typical characteristic of this mode is that lower interface occurs. The solid concentration of sediment bed changes abruptly from $\phi_0$ to $\phi_{max}$. Because of this sudden change in concentration, the settling velocity reduces to zero instantly, so the curved part of settling curve in Case 1 is insignificant (see Fig. 3 (a)). Another mode of sedimentation in Case 2 when initial solid concentration $\phi_0 = 0.2, 0.3, 0.4$, is similar to MS-IV or MS-V sedimentation described by Bustos and Concha [7,8]. A typical characteristic of this mode is that the solid concentration of sediment bed change gradually from $\phi_0$ to $\phi_{max}$, so the lower interface would not be clearly captured. The gradual increase of the solid concentration leads to the gradual decrease of particle settling velocity from $V_s$ to 0, which can be displayed by the curved part (gradual reduction of the slope) of the settling curves (see Fig. 3 (b)). In summary, by employing CFD–DEM method, two typical modes of sedimentation are simulated, and the forms of concentration profiles and characteristic lines obtained from simulations are similar to those in the theory of sedimentation. The different mode of sedimentation occurs depending on whether the Oleinik's jump entropy condition is satisfied (see Eq. (19)-(23)). The critical time when sedimentation ends of these two modes of sedimentation will be calculated in Sec. 4.

### 3.4 The segregation phenomenon of the poly-dispersed particle system

According to the equations calculating the terminal velocity of a particle falling in liquid under the effect of gravity [36], the greater the size of the particle is, the larger the particle's terminal falling velocity will be. Therefore, when poly-dispersed non-cohesive particles fall in the liquid, larger particles will settle through the small particles to the base, leaving the top layer with a high percentage of finer particles. This segregation phenomenon was observed in experiments of coarse silt settling when the initial solid concentration is rather low, e.g., the initial density is 10.7 kN/m$^3$ [37], approximately $\phi_0 = 0.03$. In our simulations, Case 3 studies the settling process of poly-dispersed particles. The particle size follows the log-normal distribution, similar to the diameter distribution of natural sands in rivers [38], and the parameters of this case are given in Tab. 1. The solver, sediFoam, employed in our simulations, has great ability to handle the situation that the particle size is larger than fluid cell [23,39], thus could accommodate wide particle size distributions of Case 3.

Figures. 7 (a) and (b) show the particle size distribution (PSD) at different snapshots when $\phi_0 = 0.05$. It indicates that when sedimentation ends, the top layer of sediment bed consists of more fine particles, and the base consists of more large particles, which means that the particle segregations occur during the settling process. Moreover, the falling velocity of the top-layer particles of the poly-dispersed system is smaller than the mono-dispersed system due to the segregation effect (see Fig. 8). The superficial layer of poly-dispersed suspension contains a higher percentage of the fine particles, leading to the small effective falling velocity. While this segregation effect reduces with the increase of the initial concentration. By comparing the PSD at different $\phi_0$ (see Figs. 7 (b) $\phi_0 = 0.05$ and (d) $\phi_0 = 0.4$), the separation of particle size is less obvious at high initial concentration. Moreover, with the growth of $\phi_0$, the velocity deviations between the poly- and mono-dispersed particles decrease (see Fig. 8). This is because when the concentration of suspension $\phi_0$ is great, there are more intensive collisions among particles which enforce small particles falling together with the big particles, so the particle segregation effect reduces. The reduction of the segregation effect is also observed in Been's experiments [37], e.g., the initial density of this experiment is 12.0 kN/m$^3$, approximately $\phi_0 = 0.13$.

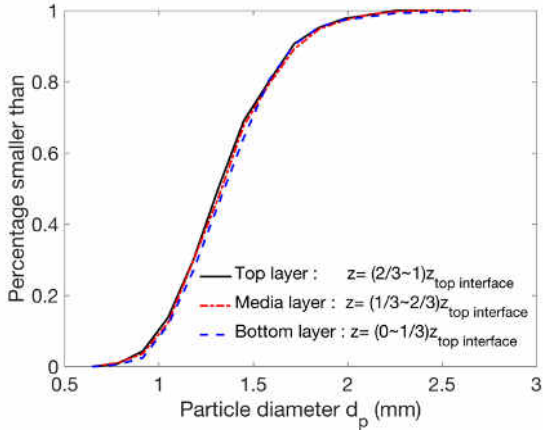
(a) $t = 0$, $\phi_0 = 0.05$

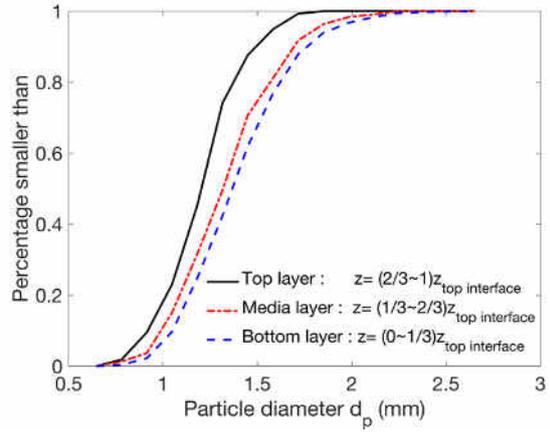
(b) At the end of simulations, $\phi_0 = 0.05$

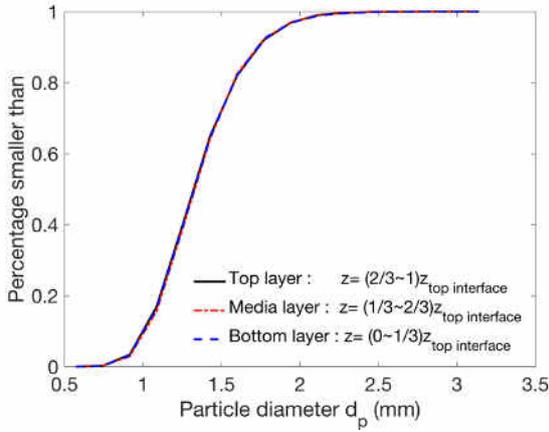
(c) $t = 0$, $\phi_0 = 0.4$

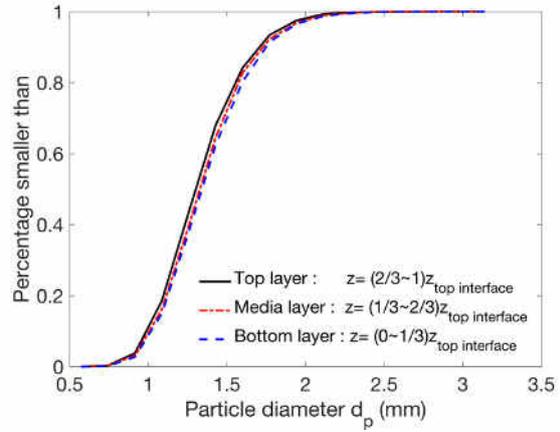
(d) At the end of simulations, $\phi_0 = 0.4$

**Fig. 7 Particle size distribution (PSD) at different snapshots of mixtures when (a) $t = 0$, $\phi_0 = 0.05$; (b) At the end of simulations, $\phi_0 = 0.05$; (c) $t = 0$, $\phi_0 = 0.4$ (d) At the end of simulations, $\phi_0 = 0.4$**

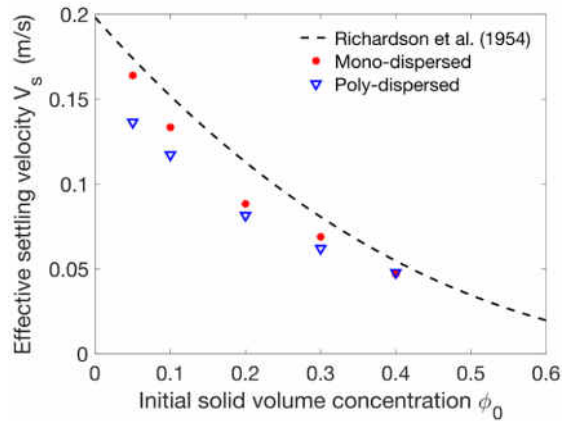

**Fig. 8 Comparison of the falling velocities of superficial suspensions between the poly-dispersed particles, the mono-dispersed particles, and those calculated by Richardson's formula. The empirical factor n in Richardson's formula is calculated by Eq. (19), of which the value is 4.65 regardless of the diameters of particles. It is because of the lateral periodic boundaries that the wall effect term $d_p/D$ is neglected.**

In fact, not only the particle segregations but also the variations of local concentration of suspensions due to the segregation, both affect the effective falling velocity. However, their individual influence on the settling velocity, and the specific formula to calculate settling velocity of the superficial poly-dispersed suspensions have not been discussed in previous research. What we observed in our simulations, is the segregation phenomenon and the falling velocity discrepancy between the poly-dispersed system and mono-dispersed system. Moreover, the segregation effect reduces with the growth of the $\phi_0$ due to the more intensive interferences in the falling of small particles by the big particles in the poly-dispersed particle system.

**3.5 The correspondence between micro-particle contact force and effective stress**

By using CFD–DEM method, the evolution of the fluid pressure and particle contact force can be obtained. The time-history of the excess pore pressure on the bottom of simulation domain is shown in Fig. 9. We find that the excess pore pressure decreases linearly, and its maximum value appears when the weight of all particles is supported by the fluid at $t = 0$. The simulation results are fitted by a straight line and the intercept of the line is 645.30 Pa (see Fig. 9). The analytical value of maximum excess pore pressure can be calculated by [19]:

$$u_{e\_max} = \phi_0 \rho_p h_{box} g + (1 - \phi_0)\rho_f g h_{box} = \phi_0 (\rho_p - \rho_f) g h_{box}, \tag{24}$$

where $h_{box}$ represents the height of the simulation domain in z-direction. Substituting the setup of Case 2 when $\phi_0 = 0.3$, the maximum analytical excess pore pressure equals to 654.88 Pa, which is in good agreement with 645.30 Pa obtained in the simulations.

The vertical profile of excess pore pressure is shown in Fig. 10. From this figure, three different parts can be distinguished along the vertical direction. The highest part is the pure water, of which the excess pore pressure is zero. The middle part is the suspension, of which the excess pore pressure increases linearly with the depth to support the submerged weight of suspended particles. The lowest part represents the sediment bed, of which the excess pore pressure does not vary with the depth. It's because once particles settle down and contact each other that the submerged weight of these deposited particles will be balanced by contact force. Therefore, the excess pore pressure keeps constant in the deposited particles, and the value of it equals to the submerged weight of all suspended particles. It is worth mentioning that the increase of the contact force stress and the dissipation of excess pore pressure are instantaneous when sand particles deposit in water, which is not the same as that during the consolidation process among cohesive particles [4].

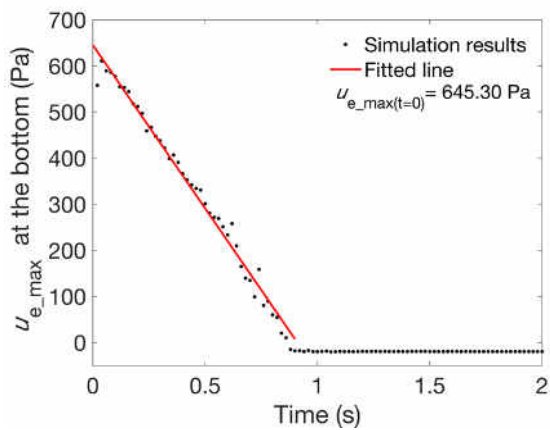 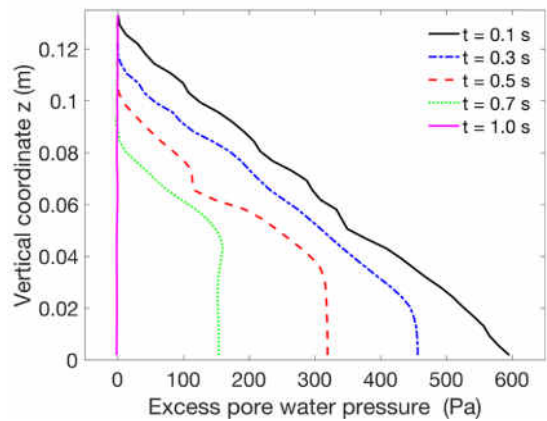

Fig. 9  Excess pore water pressure on the bottom of simulation domain when $\phi_0 = 0.3$ in Case 2. The simulation results are fitted by a straight line and the intercept of the line when $t = 0$ is 645.30 Pa

Fig. 10 The profile of excess pore fluid pressure along $z$ direction at different time when $\phi_0 = 0.3$

Then the relationship between the effective stress and the contact force exerted on particles are introduced here. The effective stress $\sigma'$, a critical concept widely used in soil mechanics, is proposed by Terzaghi [40]. The term "effective" means the calculated stress is effective in moving soil, and the "effective stress" represents the average stress carried by soil skeleton. The value of σ′ acting on soil, which is calculated by effective stress principle [40], equals to the total stress σ subtracting the pore fluid pressure, including the static fluid pressure $p_s$ and the excess pore water pressure $u_e$,

$$\sigma' = \sigma - p_s - u_e, \tag{25}$$

where σ represents the total stress of sediment at a certain height. The total stress can be calculated by summing up the total stress of three individual parts:

$$\begin{aligned}\sigma &= \sigma_{\text{sedi}} + \sigma_{\text{sus}} + \sigma_{\text{water}} \\ &= \left[\phi_{\max}\rho_p + (1-\phi_{\max})\rho_f\right]gh_{\text{sedi}} + \left[\phi_{\text{sus}}\rho_p + (1-\phi_{\text{sus}})\rho_f\right]gh_{\text{sus}} + \rho_f gh_{\text{water}},\end{aligned} \tag{26}$$

where the subscript sedi, sus, water respectively represents the part of sediment bed, suspension between two interfaces, and pure water. $\phi_{\max}$ represents the solid concentration of sediment bed. $h$ is the vertical thickness of these three parts of mixtures, and $h_{\text{sedi}}$ represent the thickness of sediment bed above a certain location. In Eq. (25), the static fluid pressure $p_s = \rho_f g(h_{\text{sedi}} + h_{\text{sus}} + h_{\text{water}})$. The excess pore fluid pressure equals to the submerged weight of suspended particles (see Fig. 9), $u_e = \phi_{\text{sus}}(\rho_p - \rho_f)gh_{\text{sus}}$. Substituting $\sigma, p_s, u_e$ into Eq. (25), we can calculate the theoretical effective stress in sediments by:

$$\sigma' = \phi_{\max}(\rho_p - \rho_f)gh_{\text{sedi}}. \tag{27}$$

In Case 1, when $\phi_0 = 0.3$, the value of $\phi_{\max}$ is 0.581 from the simulation. The theoretical effective stresses calculated by Eq. (27) at different time are shown in Fig. 11 by straight lines. Then, we define the "*mean contact stress*" by the following formula:

$$\sigma^{\text{con}} = \frac{\sum f_{p,i}^{\text{con}}}{A_c}, \tag{28}$$

where $\sum f_{p,i}^{\text{con}}$ is a sum of contact forces of every pair of particles at the same height, and $A_c$ is the cross-sectional area of computational domain. The points plotted in Fig. 11 represent the simulation results obtained by using Eq. (28). From Fig. 11, we find that there is great agreement between the "effective stress" σ′ and the "*mean contact stress*" $\sigma^{\text{con}}$. This validates the "effective stress principle" and connects the micro-scale $\sigma^{\text{con}}$ with macro-scale σ′. By employing Eq. (28), the contact forces between micro-particles can be converted to the effective stress, which helps to apply the simulation results to many practical civil engineering problems, such as the slope stability, the soil liquefaction, and many groundwater-related subsidences. Therefore, the correlation studied in this paper can enlarge the application of CFD–DEM model in the civil engineering field.

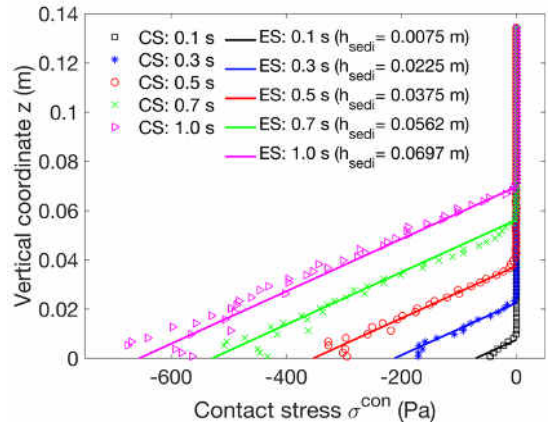

**Fig. 11 The profile of the mean contact stress (CS) $\sigma^{\text{con}}$ and the effective stress (ES) σ′ along z direction at the different time; The values of CS are obtained from simulation results and Eq. (26); Effective stress (ES) is based on the effective stress principle, calculated by Eq. (25)**

### 3.6 The solid concentration of sediment bed

After the hindered settling, particles gradually deposit and form the sediment bed. In this paper, the deposition of two different particle-fluid system is simulated, therefore, the influence of the particle properties on the solid concentration of the sediment bed $\phi_{max}$ can be investigated. Among the particle parameters, the stiffness constant and the damping constant have negligible effect on $\phi_{max}$ in our simulations and the previous researches [31,41], however, the friction coefficient $\mu$ and the initial solid concentration of suspension $\phi_0$ do influence the $\phi_{max}$, so this section discusses their effect on $\phi_{max}$ respectively.

In our simulations, the influence of friction force is considered by using the inner-particle friction coefficient $\mu$. Figure 12 shows the settling curves with different $\mu$ values ranging from 0 to 0.4. The value 0.4 is based on the experimental result for sand particles submerged in water [42]. We find that the settling curves with different $\mu$ overlap initially (see Fig. 12 ). It means that the friction force does not influence the hindered settling process and the effective settling velocity, while, the effect of friction force on the thickness and the packing density of sediment bed is non-negligible. With the increase of the friction coefficient, the height of sediment bed rises (see Fig. 12), and the concentration of sediment bed $\phi_{max}$ decreases as shown in Tab. 2. This is because the solid concentration of sediment bed $\phi_{max}$ is related to the average number of contacts per particle [43]. Furthermore, the contacts per particle depend fundamentally on the forces acting on the particle. Among these forces, the friction resists particle sliding on another, and this sliding resistance tends to be larger with greater friction coefficients. Thus, when coarser particles drop on sediment bed, they are less likely to slide and more likely to form more loose-packing structures. As shown in Tab. 2, when $\mu$ in Case 1 is 0.4, $\phi_{max}$ is 0.58, which is in the range of the experimental packing density of sand particle 0.58~0.6 [44]. Therefore, the friction coefficient of Case 1 is determined as 0.4. For the Case 2, there is insufficient experimental value about friction coefficient of ballotini deposited in the silicon. When the friction coefficient $\mu$ is taken from [0, 0.1], the packing density of spherical particles obtained in the simulations is consistent with the experimental results [45]. When $\mu$ in Case 2 is 0.02, $\phi_{max}$ is 0.624, very close to the value 0.625, which is the theoretical pack density of spherical particles [45].

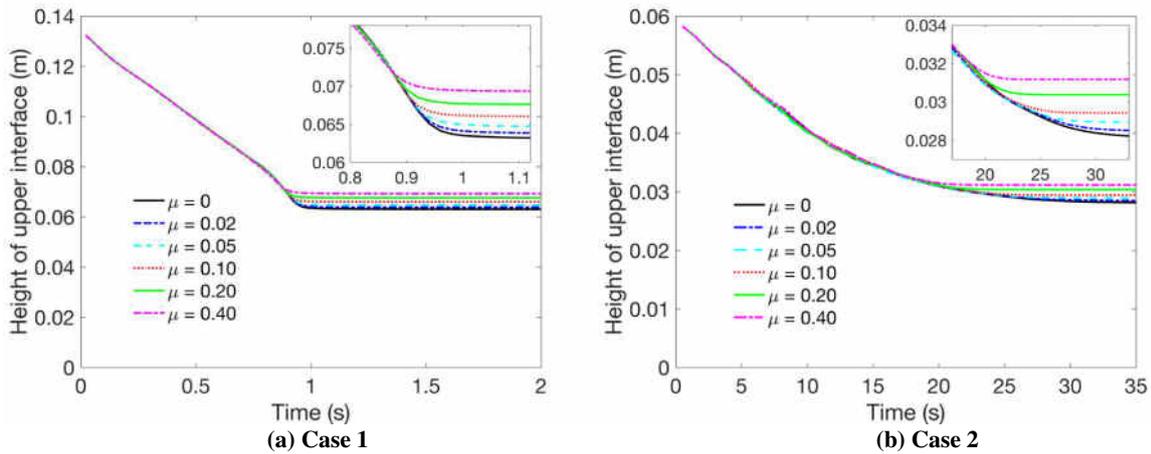

(a) Case 1      (b) Case 2

Fig. 12 The settling curves of (a) Case 1 and (b) Case 2 with different frictional coefficient $\mu$ when $\phi_0 = 0.3$

Tab. 2 The solid concentration of sediment bed with different friction coefficients when $\phi_0 = 0.3$

| Friction coefficients | 0 | 0.02 | 0.05 | 0.1 | 0.2 | 0.4 |
|---|---|---|---|---|---|---|
| Case1 | 0.6354 | 0.6271 | 0.6193 | 0.6088 | 0.5949 | 0.5809 |
| Case2 | 0.6316 | 0.6244 | 0.6134 | 0.6038 | 0.5839 | 0.5689 |

With friction coefficient $\mu = 0.4$ and $\mu = 0$ in Case 1, we investigate the influence of the initial solid concentration of suspension $\phi_0$ on the value of $\phi_{max}$. The vertical concentration profiles of sediment bed when the sedimentation ends are shown in Fig. 13, and the values of $\phi_{max}$ on different conditions are shown in Fig. 14. These figures show that with the increase of $\phi_0$, the $\phi_{max}$ increase. It is because the particles at the same z-coordinate will suffer the greater downward effective forces which balance a part of friction. The sliding resistance of friction tends to be decreased, so the particles are prone to form a more close-packing structure. While, when particles are totally smooth ($\mu = 0$), the solid concentrations of sediment bed $\phi_{max}$ do not change significantly with the increase of $\phi_0$. This indicates that without the effect of friction, smooth particles (randomly distributed in suspension) will fall freely and pack in a fixed manner which is the most prone to happen regardless of the initial concentration $\phi_0$.

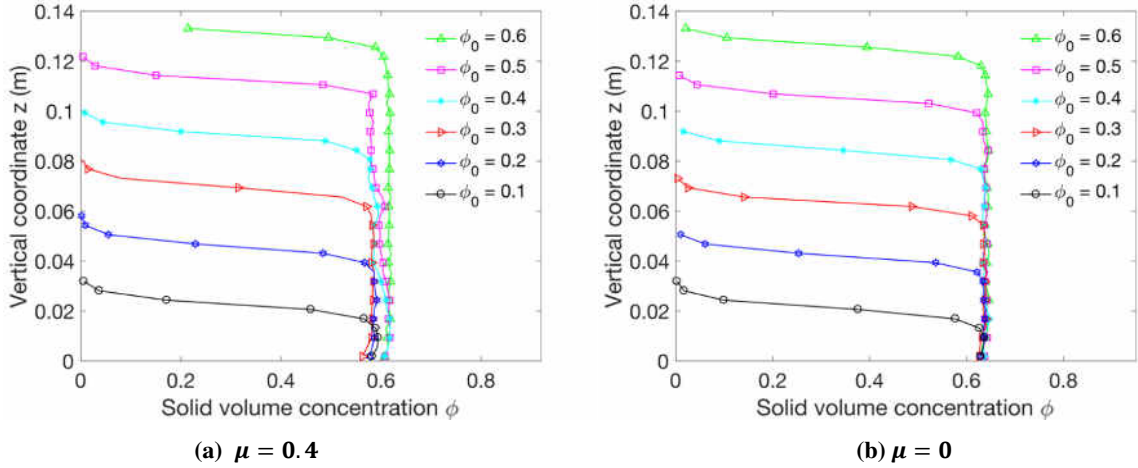

(a) $\mu = 0.4$  (b) $\mu = 0$

**Fig. 13 The vertical concentration profiles of sediment bed at different initial concentrations of Case 1 when (a) $\mu = 0.4$ and (b) $\mu = 0$**

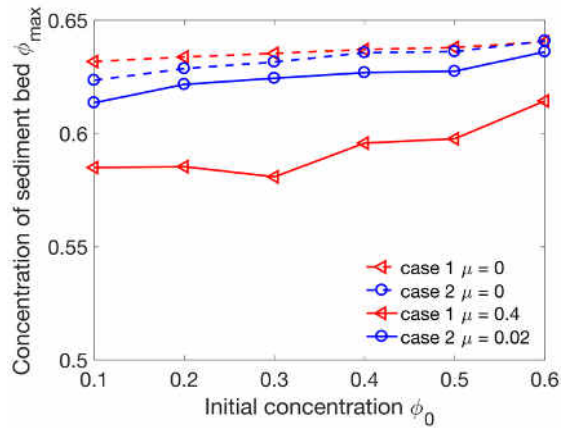

**Fig. 14 The variation of averaged solid concentration of sediment bed $\phi_{max}$ on different conditions**

## 4  Discussion

At the final state of any modes of sedimentation, there are two regions in the mixtures: a sediment bed with $\phi_{max}$ and a layer of clear water on top ($\phi = 0$). The time when this final state arises is called critical time, denoted by $T_c$ here. This time could not only be measured from the experiments or the simulations but also be theoretically estimated by mathematical modeling. However, the study on the latter method is lacking. One of the reasons is that mathematical analysis is based on the specific flux–concentration function $S(\phi)$ and the maximum concentration $\phi_{max}$, which are not both commonly determined. In this part, we will discuss this mathematical method, and the calculation results will be compared with the simulation results.

The function $S(\phi)$ we utilized here is the Richardson's empirical falling velocity formula. The $S(\phi) - \phi$ curve of Case 1 and Case 2 is shown in Fig. 15. It is worth mentioning that the function $S(\phi)$ and the $S(\phi) - \phi$ curve should be revised before being used to calculate the critical time. Based on the Richardson's formula, the settling velocity $V_s$ is zero when $\phi = 1$ corresponding to $S(1) = 0$ (see dashed lines in Fig. 15). But the reality is that when the solid concentration increases to its maximum packing value $\phi_{\max}$ ($\phi_{\max} < 1$), the settling velocity $V_s$ has decreased to zero [5]. Therefore, Fitch [46] proposed that the $S(\phi) - \phi$ curve should satisfy this condition and should be cut by a vertical line at the point ($\phi_{\max}$, 0) (see solid lines in Fig. 15). Furthermore, Bustos and Concha [7,8] proposed that this form of $S(\phi)$ is similar to that with two inflection points. However, they did not individually interpret this revised $S(\phi) - \phi$ curve and describe how to obtain the critical time. Here, the critical time of this practical sedimentation will be calculated based on the sedimentation theory [5] and the mathematical analyzing [7,8].

For the sedimentation of Case 1, the 'entropy condition' is fulfilled (see Eq. (9)) at any initial concentration $\phi_0$, the MS-I sedimentation occurs (mentioned in Sec. 3.3). When the sedimentation ends, the upward suspension–sediment interface will meet with the downward liquid–suspension interface at one point which corresponds to the critical time (see Fig. 5 (b)). The upward velocity of suspension–sediment interface $U_{\text{shock}}$ can be calculated by Eq. (8), the concentration of two points locating just above and below the interface is $\phi_0$ and $\phi_{\max}$, respectively. Also, in $S(\phi) - \phi$ curve, the value of $U_{\text{shock}}$ corresponds to the slope of the chord connecting the point ($\phi_{\max}$, 0) and point ($\phi_0, S(\phi_0)$) (see red dotted lines in Fig. 15 (a)). The z-coordinate of the terminal height of sediment bed $H_{\text{sedi}}$ can be obtained by $H_0 \phi_0 / \phi_{\max}$, where $H_0$ represents the initial height of the suspension. Therefore, the critical time $T_c$ when sedimentation ends of MS-I sedimentation could be calculated by:

$$T_c = H_{\text{sedi}} / U_{\text{shock}} = \frac{\phi_0 H_0}{\phi_{\max} U_{\text{shock}}} = \frac{\phi_0 H_0 (\phi_0 - \phi_{\max})}{\phi_{\max}(S(\phi_0) - S(\phi_{\max}))}. \tag{29}$$

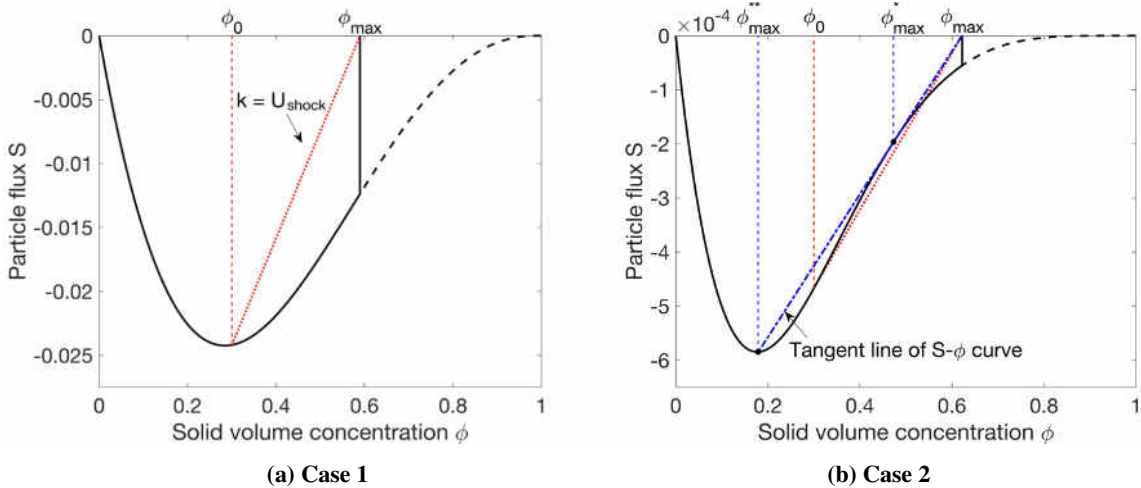

(a) Case 1     (b) Case 2

**Fig. 15** $S(\phi) - \phi$ curve of (a) Case 1 and (b) Case 2; The solid line is the $S(\phi) - \phi$ curve calculated by Richardson's formula and is cut by a vertical line through point ($\phi_{\max}$, 0); The red dotted lines denote the chord of $S(\phi) - \phi$ curves; The blue dash-dot line in (b) Case 2 represents the tangent line of $S(\phi) - \phi$ curve through the point ($\phi_{\max}$,0), and the black points ($\phi_{\max}^{**}, S(\phi_{\max}^{**})$) and ($\phi_{\max}^{*}, S(\phi_{\max}^{*})$) are the intersection points

For the sedimentations of Case 2, different modes of sedimentation occur at different $\phi_0$. Based on Bustos and Concha [7,8], when initial concentration $\phi_0$ is in the range [$\phi_{\max}^{**}, \phi_{\max}^{*}$] ($\phi_{\max}^{**}, \phi_{\max}^{*}$ represent the intersection points of the tangent line of $S(\phi) - \phi$ curve through ($\phi_{\max}$, 0) and the $S(\phi) - \phi$ curve as shown in Fig. 15 (b)), the chord of $S(\phi) - \phi$ curve (red dotted line in Fig. 15 (b)) connecting the point ($\phi_{\max}$, 0) and the point ($\phi_0, S(\phi_0)$) will intersect curve itself. The "entropy condition" is not fulfilled anymore, and the suspension–sediment interface (shock) can't exist stability. At these $\phi_0$, MS-IV or MS-V sedimentation occurs, the liquid–suspension interface has many intersection points with the lower rarefaction wave (see Fig. 6 (b)), so the critical time when sedimentation ends can not be calculated by Eq. (29). Based on Bustos' analyzing, the critical time of this mode of sedimentation can be calculated by:

$$T_c = H_{\text{sedi}}/S'(\phi^*_{\max}) = \frac{\phi_0 H_0}{\phi_{\max} S'(\phi^*_{\max})}, \qquad (30)$$

where $S'(\phi^*_{\max})$ is the slope of the tangent line of the $S(\phi) - \phi$ curve through the point $(\phi_{\max}, 0)$ (see Fig. 15 (b)). When initial concentration $\phi_0 < \phi^{**}_{\max}$ or $\phi_0 > \phi^*_{\max}$, the MS-I will happen. The velocity of lower shock can be calculated by Eq. (8), and the critical time could be obtained by Eq. (29).

The critical time $T_c$ of Case 1 and Case 2 calculated by Eq. (29) or Eq. (30) are shown in Tab. 4. For the sedimentation of Case 2, when $\phi_0 = 0.2, 0.3$ and $0.4$, the $\phi_{\max}$ is 0.622, 0.624, 0.627, respectively. The concentration of intersection point $\phi^{**}_{\max}$ obtained from $S(\phi) - \phi$ curve is in the range [0.169, 0.179], and $\phi^*_{\max}$ is in the range [0.473, 0.483]. At these $\phi_0$, it satisfies the conditions that $\phi^{**}_{\max} < \phi_0 < \phi^*_{\max}$, so the critical time is calculated by Eq. (30). When the initial concentration is 0.1, 0.5, 0.6, the MS-I occurs, so the critical time $T_c$ is calculated by Eq. (29). Moreover, we compare the calculation results of $T_c$ with the simulation results as shown in Tab. 4 and Fig. 16. The time when the sedimentation ends in the simulations is obtained when the variation of the sediment height is negligibly small, denoted by $T_r$.

**Tab. 3 Calculation results of critical time in Case 1**

| $\phi_0$ | $\phi_{\max}$ | $\sigma$ | $H_{\text{sedi}}$ | $T_c$ | $T_r$ | Deviation |
|---|---|---|---|---|---|---|
| 0.1 | 0.585 | 0.0313 | 0.0231 | 0.74 | 0.88 | 0.163 |
| 0.2 | 0.585 | 0.0586 | 0.0461 | 0.79 | 0.94 | 0.163 |
| 0.3 | 0.581 | 0.0862 | 0.0697 | 0.81 | 0.94 | 0.139 |
| 0.4 | 0.596 | 0.1118 | 0.0906 | 0.81 | 0.88 | 0.079 |
| 0.5 | 0.598 | 0.1770 | 0.1129 | 0.64 | 0.64 | 0.003 |
| 0.6 | 0.614 | 1.1827 | 0.1328 | 0.11 | 0.12 | 0.064 |

**Tab. 4 Calculation results of critical time in Case 2**

| $\phi_0$ | $\phi_{\max}$ | $\sigma$ | $H_{\text{sedi}}$ | $T_c$ | $T_r$ | Deviation |
|---|---|---|---|---|---|---|
| 0.1 | 0.614 | 0.000975 | 0.00978 | 10.0 | 11.0 | 0.088 |
| 0.2 | 0.622 | 0.001322 | 0.01930 | 14.6 | 18.0 | 0.189 |
| 0.3 | 0.624 | 0.001303 | 0.02889 | 22.2 | 27.0 | 0.179 |
| 0.4 | 0.627 | 0.001274 | 0.03828 | 30.1 | 35.5 | 0.153 |
| 0.5 | 0.628 | 0.001276 | 0.04780 | 37.4 | 42.5 | 0.119 |
| 0.6 | 0.636 | 0.001915 | 0.05659 | 29.5 | 27.5 | -0.074 |

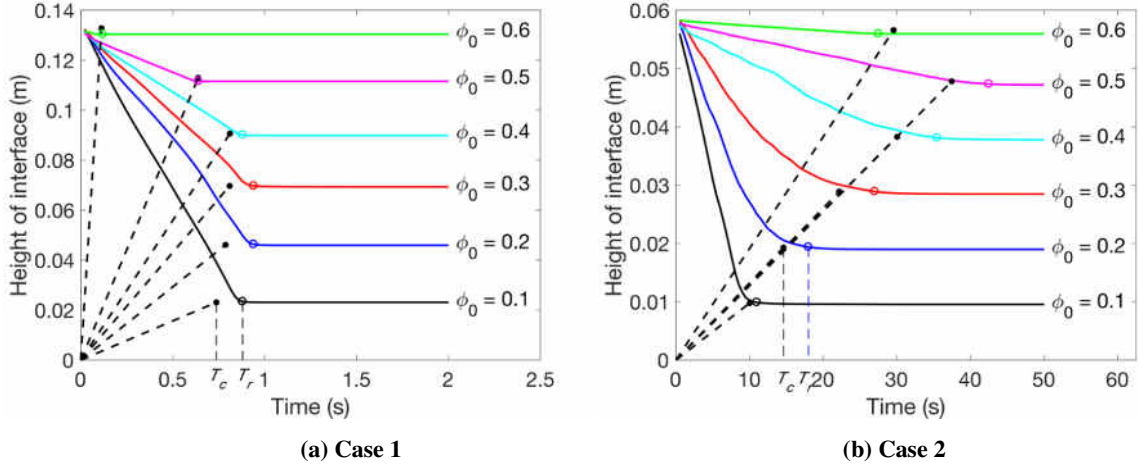

(a) Case 1  (b) Case 2

Fig. 16 The comparison between the calculation $T_c$ and simulations results $T_r$ when sedimentation ends in (a) Case 1 and (b) Case 2;

From above calculation and comparison, the mutual deviations between the $T_c$ and $T_r$ amount to 0.101, 0.133 (average), and 0.163, 0.189 (maximum). Although the deviation is not very small, while, it demonstrates that the formulae proposed above and the simulations could both be applied to obtain the critical time. The precondition of proposed formula is that when concentration gets to the maximum value $\phi_{max}$, the settling velocity should be zero. Therefore, when calculating the critical time, this precondition, which did not draw our much attention before, should be satisfied. Moreover, above calculations provide a simple approach to approximately estimate the time when sedimentation end. Once we know the maximum concentration of sediment bed $\phi_{max}$ from the simulations or the experiments, the revised $S(\phi) - \phi$ curve can be plotted with an empirical falling velocity formula, then the critical time when sedimentation ends can be calculated by Eq. (29) or Eq. (30).

## 5 Conclusions

In this paper, the hindered settling and the deposition process was simulated by CFD–DEM method. By employing this method, the velocity–concentration relationship, the micro-particle contact force during the sedimentation process, and the solid concentration of sediment bed were investigated here. Moreover, the theory of hindered settling was reviewed, discussed and compared with the simulation results. The concluding remarks are as follows.

The velocity–concentration relationship is validated by two separate simulations: the settling of (1) sand in water and (2) ballotini in silicon. The agreement demonstrates the ability of the CFD–DEM model to simulate the settling velocity of the different particle-suspension systems at a wide range of initial concentration $\phi_0$ = [0.05, 0.6]. The two typical modes of sedimentation are observed by the concentration profiles and the characteristic lines. Moreover, in the settling process of the poly-dispersed particle system, the segregation phenomenon happens. The smaller particles have lower effective settling velocity, so small particles are left to the upper granular suspension. The existence of the small particles decreases the effective falling velocity of poly-dispersed particle system. However, with the increase of initial suspension concentration, small particles are enforced falling together with the big particles; hence, the segregation effect reduces, and the falling velocity gap between the poly-dispersed particles and mono-dispersed particles reduces.

During the sedimentation process, the variation of fluid pressure and the micro-particle contact force show that the submerged weight of the suspended particle is balanced by excess pore pressure, and the submerged weight of the sedimentary particle is balanced by contact force. In addition, we demonstrate the effective stress principle from the view of the contact force between particles. The micro-particle contact force can be converted to the effective stress by Eq. (28). This can help to enlarge the application of CFD–DEM model in more and more practical civil engineering projects.

The solid concentration of sediment bed $\phi_{max}$ is another quantity of interest. We find that the friction force among particles plays an important role in the concentration of sediment bed. The friction force resists the sliding of particles, so the particles are prone to form a more loose-packing structure with greater friction force. Therefore, the greater friction coefficient and the lower initial concentration, which both corresponds to the greater effective friction force, will lead to the decrease of the solid concentration of sediment bed $\phi_{max}$. However, when particles are smooth, the sedimentary particles will arrange themselves in a fixed manner regardless of the initial concentration, so $\phi_{max}$ nearly keeps constant at different $\phi_0$. In the last, a simple method to calculate the critical time when sedimentation ends is presented. It's worth noting that calculations should be on the condition that $S(\phi) - \phi$ curve passes through the point $(\phi_{max}, 0)$.

However, our simulations still have a few limitations. In this study, the particles are both simulated as a sphere, and the irregularity of realistic particles are not considered. From the experiments of realistic sand grains settling, the empirical factor n is greater which means the settling velocity is lower because of the irregularities of particles [47]. Therefore, in further research on sedimentation process, the irregularity of particle will be investigated in simulations. Moreover, the complicated non-contact interaction force between particles, such as van der Waals force and the electrostatic force, will be considered in our CFD–DEM model to simulate the cohesive particle system (such as mud/silt in civil engineering).

## Acknowledgement

The research described here was funded by the *National key research and development program funded by Ministry of Science and Technology*, grant number is 2016YFC0800207; and the *Projects of International Cooperation and Exchanges NSFC* (National Natural Science Foundation of China), grant number is 51620105008.

**Conflict of interest**: The authors declare that they have no conflict of interest.